\documentclass{aa}
\usepackage{epsfig}
\begin{document}

\thesaurus{11.05.2;\hfill  published in {\it A\&A}, {\bf 360}, L5 (2000)}
\title{Confidence levels of evolutionary synthesis models}

\author{M.~Cervi\~no\inst{1,2}\thanks{ESA posdoctoral fellow.}
        \and V. Luridiana\inst{3} 
        \and F.J. Castander\inst{1}}
\institute{Observatoire Midi-Pyr\'en\'ees, 14, avenue Edouard Belin, 
               31400 Toulouse, France 
           \and Centre d'Etude Spatiale des Rayonnements, CNRS/UPS,
B.P.~4346, 31028 Toulouse Cedex 4, France 
           \and Instituto de Astronom\'\i a, UNAM, Apdo. Postal
           70-264, 04510 M\'exico D.F., M\'exico}

\offprints{Miguel Cervi\~no}
\mail{mcervino@obs-mip.fr}
\date{Received 8 June 2000/ Accepted 30 June 2000}
\authorrunning{M.~Cervi\~no et al.}
\titlerunning{Confidence levels of synthesis models}
\maketitle

\begin{abstract}
  
  The stochastic nature of the IMF in young stellar clusters implies that
  clusters of the same mass and age do not present the same unique values
  of their observed parameters. Instead they follow a distribution. We
  address the study of such distributions, parameterised in terms of their
  confidence limits, in evolutionary synthesis models. These confidence
  limits can be understood as the inherent uncertainties of the synthesis
  models. Here we concentrate on some parameters such as EW(H$\beta$) in
  emission.  For instance, we show that for a cluster where 10$^5$
  M$_\odot$ have been transformed into stars, the dispersion of
  EW(H$\beta$) is about 18\% within the 90\% confidence levels at ages
  between 3.5 and 5 Myrs.

\end{abstract}

\section{Introduction}

Since the development of evolutionary synthesis models by
\nocite{Tin80}Tinsley (1980) several groups have developed models trying
to reproduce observable quantities in systems in which the stellar
content is not resolved. These models provide a powerful tool to study
galactic and extragalactic H{\sc ii} regions. Several types of models can
be found in the literature. For instance, \nocite{Leietal96}Leitherer et
al. (1996) give an extensive and comprehensive review on them.

The variety of external inputs used in the models (evolutionary tracks,
stellar atmosphere libraries, etc) and the treatment of these input
parameters, (for example, the interpolation methods), lead to slightly
different model outputs from different authors, although ``grosso modo''
all of them produce similar results. One way to check the applicability of
these models is to compare the results using different inputs. In this
direction we highlight the work of
\nocite{Bru2000}Bruzual (2000) in which the author presents an extensive 
study on how the external inputs to the models influence the output
results. 

These problems can be exacerbated when modelling systems with small number
of stars given that, in general, synthesis models use continuous functions
that do not reproduce exactly the discontinuous nature of star formation,
especially in systems with low number of stars. In this letter we address
this issue by studying the intrinsic variations of the evolutionary
synthesis models outputs due to the deviations of the analytical Initial
Mass Function (IMF) when modelling H{\sc ii} regions.

Here, we study the importance of such fluctuations in some typical
observables of extragalactic H{\sc ii} regions. We characterise our results
showing confidence levels of the distribution of a given parameter as a
function of age and system mass. We focus on the study of the H$\beta$
equivalent width in emission (EW(H$\beta$), that is extensively used as an
age indicator of star forming regions.  The results for the number of
ionizing photons, $Q(H^0)$, and the ratio of the blue Wolf-Rayet (WR) bump
luminosity to the H$\beta$ luminosity are also presented. We differ the
study of other quantities using different metallicities, IMF slopes, star
formation regimes and evolutionary tracks to a future paper.

\section{Synthesis code and the Initial Mass Function}

The study of the IMF has been broadly covered in the astronomical
literature. See Scalo (1986) for a comprehensive review. We define the IMF
in the following way:

\begin{equation}
\Phi(m)=\frac{dN}{dm}=A m^{-\alpha}
\end{equation}

\noindent where $\alpha$ is the IMF slope, and $A$ is a normalisation
factor.  This function gives us the number of stars in a given mass range.
The widely used Salpeter IMF corresponds to $\alpha = 2.35$ with this
definition. The total mass of the system will be

\begin{equation}
M=\int_{m_{\rm low}}^{m_{\rm up}} m \Phi(m) dm =\int_{m_{\rm low}}^{m_{\rm up}} m dN
\end{equation}

In an evolutionary synthesis code, the amount of stars produced is usually
generated binning the IMF (or using Montecarlo simulations). The evolution
of each bin (or star) is followed along the corresponding evolutionary
track and the final result is computed from the overall evolution of all
the stars (or masses) considered. In such a procedure, variations in the
IMF affect directly the output results. The IMF and the synthesised
parameters are usually normalised to the total mass transformed into stars
in the synthetic cluster. The comparison with real data is always carried
out comparing an observed quantity with the corresponding one obtained with
the code. It is assumed that the synthesised quantity is a good
approximation to the theoretical analytical value and usually no further
considerations are taken into account when comparing data and model
outputs.

We have used an updated version of the evolutionary synthesis code of
\nocite{CMH94}Cervi\~no \& Mas-Hesse (1994), that uses Montecarlo
realisations of the IMF to produce star forming regions of a given mass. We
have also used the code to generate analytical simulations of clusters of
the same mass, where analytical means a mass binning approximation of the
IMF.  We have tested that the analytical and the Montecarlo outputs show
consistent results. The dispersion between them is lower than 1.5\% when a
large number of stars is used ($5\times10^5$ stars, i.e. $3\times 10^6$
M$_\odot$ transformed into stars in the mass range 2 -- 120 M$_\odot$). In
this version of the code we use the solar metallicity evolutionary tracks
of \nocite{Mey94}Meynet et al. (1994).  The analytical results have also
been compared with the predictions of the code of
\nocite{Leietal99}Leitherer et al. (1999), that uses the same evolutionary
tracks.  The results obtained with both codes are in general agreement.

We have performed 600 Montecarlo realisations of a cluster in which 10$^3$
M$_\odot$ are transformed into stars in the mass range 2 -- 120 M$_\odot$;
400 realisations of a cluster of 10$^4$ M$_\odot$ and 200 of a cluster with
10$^5$ M$_\odot$.  The synthesis code is the same in all Montecarlo
realisations, the only difference being the stochasticity of the Montecarlo
process which reflects the fluctuations of the actual mass distribution
with respect to the analytical continuous IMF.

\section{Distribution of observed parameters: Confidence levels}

In this section we discuss the resulting distributions of the observable
parameters obtained from the different Montecarlo runs for the masses
mentioned above. We note that these distribution are in general
non-Gaussian. The confidence levels presented hereafter are computed
integrating the resulting distributions to contain the 90\% (68\%) central
values of the given parameter. We refer to these values of the parameters
as the 90\% (68\%) confidence levels of the resulting parameter
distribution obtained from the Montecarlo realisations.

\begin{figure}
\epsfig{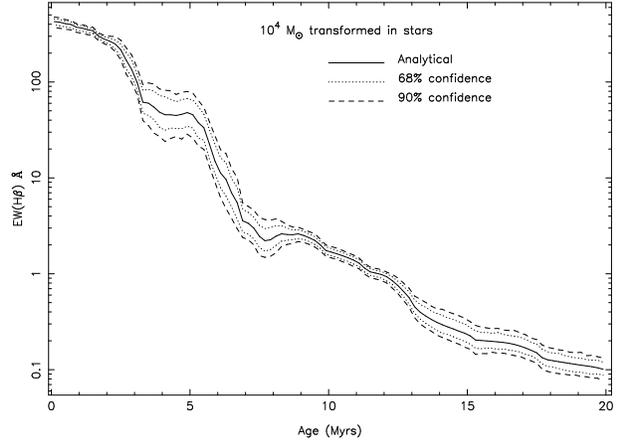}
\caption[]{90\% and 68\% Confidence levels of the EW(H$\beta$) 
distribution for the simulations with $10^4$ M$_\odot$ transformed into
stars.}
\label{fig1}
\end{figure}

Fig. \ref{fig1}. shows the 90\% and 68\% (equivalent to 1$\sigma$)
confidence levels for EW(H$\beta$) resulting from the simulations in which
$10^4$ M$_\odot$ are transformed into stars. This is the mass range in
which H{\sc ii} regions with L(H$\alpha$) $\approx$ $10^{39}$ erg s$^{-1}$
lie. This value of luminosity of H$\alpha$ is usually used as the boundary
between giant and normal H{\sc ii} regions.

It can be seen from the figure that the confidence levels change with time.
The largest spread in EW(H$\beta$) is obtained between 3 and 9 Myrs (see
also fig. \ref{fig3}) when the fluctuations in the ionizing flux and the
continuum can lead to variations of the order of 20\% at the 1$\sigma$
level. The dispersion is reduced in the age range between 9 and 13 Myrs and
increases again to an almost constant level for older ages.
 
\begin{figure}
\epsfig{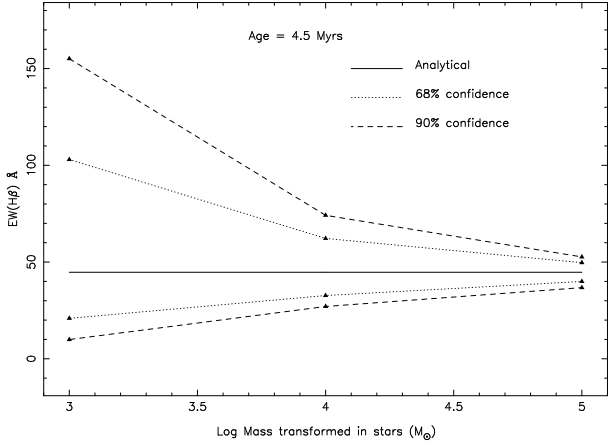}
\caption[]{90\% and 68\% Confidence levels of the EW(H$\beta$) 
  distribution for simulations of 4.5 Myrs as a function of mass
  transformed into stars.}
\label{fig2}
\end{figure}

\begin{figure}
\epsfig{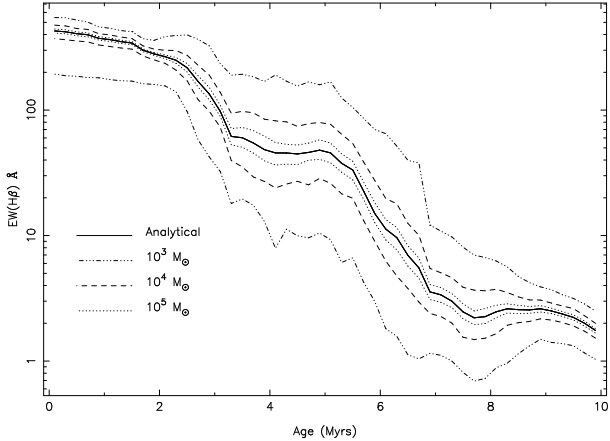}
\caption[]{Evolution of the 90\% confidence level of the EW(H$\beta$) 
distribution for simulations with different amount of gas transformed into
stars.}
\label{fig3}
\end{figure}

The mass dependence of such fluctuations can be seen in Fig. \ref{fig2}, in
which the 90\% and 68\% confidence levels are displayed for a fixed age of
4.5 Myr since the onset of the burst as a function of the mass transformed
into stars. The figure shows clearly that the dispersion is not symmetric
around the analytical value. The lower confidence level shows an almost
linear correlation with the logarithm of the mass transformed into
stars. The upper level does not show this linear behavior. An important
implication regards the different interpretations of observational data
when compared to models assuming that they provide an exact value. As can
be seen from the figures, the 90\% or 68\% confidence levels of the
parameter distributions are uncomfortably large, especially for low masses.
Due to this behaviour of the EW(H$\beta$) evolution, the age determination
between 3 and 5 Myrs are not well defined at the 90\% confidence level.
This situation may be improved using more diagnostic observables to
estimate the age since the onset of the burst.

Fig. \ref{fig3} presents the evolution with age of the 90\% confidence
level of the EW(H$\beta$) distribution for different amount of mass
transformed into stars. This figure demonstrates the importance of taking
into account the confidence levels of an observable parameter for a given
mass range.  For instance, for a 10$^5$ M$_\odot$ cluster, a value of the
EW(H$\beta$) equal to 50\AA, is compatible with any age between 3.5 and 5
Myrs within the 90\% confidence levels.

\begin{figure}
\epsfig{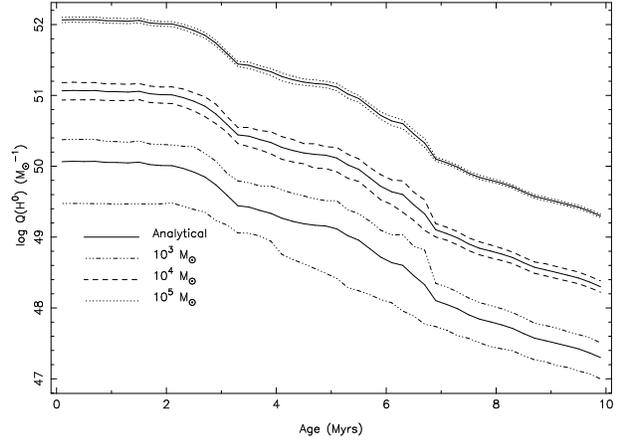}
\caption[]{Evolution of the 90\% confidence level of the $Q(H^0)$ 
distribution for simulations with different amount of gas transformed into
stars.}
\label{fig4}
\end{figure}

In Fig. \ref{fig4} the evolution of the ionizing flux, $Q(H^0)$, is
displayed against the mass transformed into stars. For a 10$^3$ M$_\odot$
system the fluctuations allow values that differ by an order of magnitude
for a given age.  The 90\% confidence level values for the distribution of
this observable show a relatively broad distribution up to 7 Myrs.
Comparing this figure with the previous one, it is possible to figure out
that the the scatter in the continuum flux around 4861 {\AA}, $L_c({\rm
H}\beta)$, is the driving parameter producing the time evolution of the
EW(H$\beta$) distribution. We have checked this point with the actual
values and confirmed it to be the case. As a consistency check, we have
also corroborated that the dispersion in $L_c(H\beta)$ combined with
the dispersion in $Q(H^0)$ approximately reproduce the dispersion in
EW(H$\beta$).

\begin{figure}
\epsfig{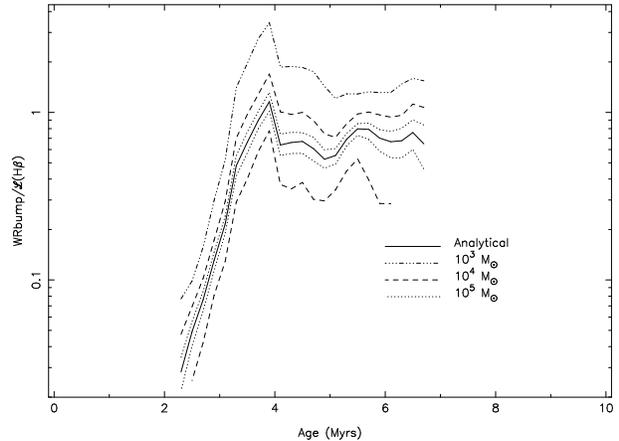}
\caption[]{Evolution of the 90\% confidence level of the 
  $L({\rm WRbump})/L(H\beta)$ distribution for simulations with
  different amount of gas transformed into stars.}
\label{fig5}
\end{figure}

Fig. \ref{fig5} shows the evolution of the ratio of the Wolf-Rayet bump and
H$\beta$ luminosities, $L({\rm WRbump})/L(H\beta)$, presenting the
90\% confidence levels of its distribution. We would like to point out that
the analytical value, close to 1, is 2.3 times higher that the one obtained
in \nocite{CMH94}Cervi\~no \& Mas-Hesse (1994). We have checked this value
and confirm that it is the result of using different evolutionary tracks
with enhanced mass loss rates. It is important to note that at the 90\%
confidence level, for a 10$^3$ M$_\odot$ cluster, values consistent with no
detection of WR stars are allowed. The distribution at 1$\sigma$ level is
also compatible with such non detection.

\begin{figure}
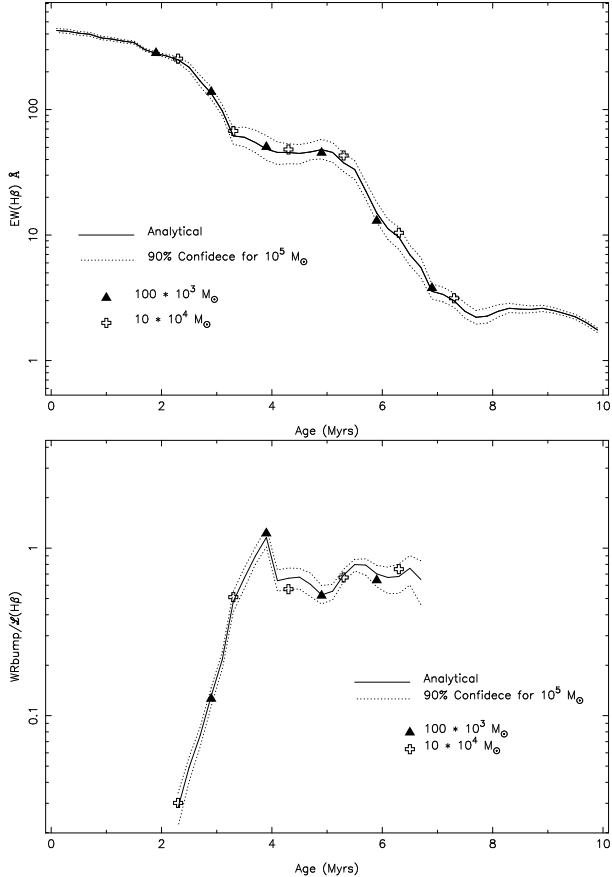

\epsfig{file=cf081fig6.ps, angle=270,  width=8cm}\\
\epsfig{file=cf081fig7.ps, angle=270,  width=8cm}
\caption[]{Comparison of the values of EW(H$\beta$) and 
$L({\rm WRbump})/L(H\beta)$ obtained for multiply-composed 10$^5$
M$_\odot$ systems (triangles and crosses) with the analytical values and
90\% confidence level for ``single'' 10$^5$ M$_\odot$ systems. (see text)}
\label{fig6}
\end{figure}

This study opens an additional question. If stochasticity of the stellar
mass distribution of low mass clusters imply such dispersions in the
distribution of observable parameters, what would its effect be in a system
composed of several smaller groups?  In order to address this question, we
have simulated multiply-composed 10$^5$ M$_\odot$ clusters by adding
randomly 6 sets of 100 clusters with 10$^3$ M$_\odot$ and 6 sets of 10
clusters 10$^4$ M$_\odot$ to obtain the observables for a variety of ages.
We have plotted the results for the EW(H$\beta$) and the $L({\rm
WRbump})/L(H\beta)$ together with the analytical values and the 90\%
confidence level of the distributions for a 10$^5$ M$_\odot$ cluster in
Fig. \ref{fig6}.

The figure shows that all the simulated points fall within the 90\%
confidence level for the 10$^5$ M$_\odot$ simulation. However, this is not
the case for the 68\% confidence level. In conclusion, the distribution of
the observed parameters due to the stochasticity of the star formation
process shrinks towards the analytical values of evolutionary synthesis
models as larger masses are considered more or less independently of
whether these systems are monolithic unique systems or composed by several
coeval subgroups. We would like to stress that our previous simulations
assumed the same age for the subgroups that form a system. We have not
addressed the consequences of different ages for the different subgroups.

As a final note we would like to point out that the total masses used here
have been computed using stellar masses in the range from 2 to 120
M$_\odot$, which are the ones relevant for the observable parameters
discussed. The corresponding total masses should be corrected by
multiplying by a factor 1.36 for stellar masses in the range from 1 to 120
M$_\odot$; multiplying by 1.5 for the mass range 0.8 to 120 M$_\odot$ and
multiplying by 3.75 for the mass range 0.08 to 120 M$_\odot$. In all the
cases a power law IMF with the Salpeter exponent has been assumed.

\section{Conclusions}

The analysis presented in this letter allow us to draw the following
conclusions:

\begin{itemize}
\item The intrinsic deviations between analytical and stochastic IMF 
  must be taken into account when comparing synthesis models with
  observational data. The inherent uncertainties depend on the amount of gas
  transformed into stars.
\item These deviations are independent of the synthesis code and represent
  a lower limit to the total uncertainties. The total deviations will
  depend on other parameters like the star formation history, the input
  ingredients, the numerical approximations, etc.
\item The deviations depend on the evolutionary tracks used. 
\item The widths of the parameter distributions compared to the analytical 
  values are proportional to the mass transformed into stars in the stellar
  cluster or groups of clusters. They also depend on the IMF slope and the
evolutionary status of the cluster.
\item The specific distribution of the deviation from the analytical value 
  varies from observable to observable. As an extreme example, the 68\%
  confidence level of the $L({\rm WRbump})/L(H\beta)$ ratio is compatible with
no
  detection of WR stars.
\end{itemize}

\begin{acknowledgements}
  We want to acknowledge useful discussion with M. G\'omez-Flechoso, J.
  Kn\"odlseder, J.F. G\'omez, J.M. Mas-Hesse and D. Schaerer. VL also
  acknowledges the Observatoire Midi-Pyr\'en\'ees for providing facilities
  to conduct this study.
\end{acknowledgements}


\end{document}